\documentclass[twocolumn]{aastex631}
\usepackage{mathrsfs}
\usepackage{xspace}

\newcommand\xmm{\textit{XMM-Newton}\xspace}
\newcommand\hst{\textit{HST}\xspace}

\newcommand\Rin{$R_{\rm in}$\xspace}
\newcommand\Rins{$R_{\rm in}^{*}$\xspace}
\newcommand\Tp{$T_{\rm p}$\xspace}
\newcommand\Rratio{$R_{\rm out}/R_{\rm in}$\xspace}

\newcommand\Rout{$R_{\rm out}$\xspace}

\newcommand\Mbh{$M_{ \rm BH}$\xspace}
\newcommand\ergs{erg s$^{-1}$\xspace}
\newcommand\msun{$M_\odot$\xspace}
\newcommand\target{eRO-QPE2\xspace}
\usepackage{threeparttable}

\begin{document}
\title{Time-resolved Hubble Space Telescope UV observations of an X-ray quasi-periodic eruption source}
\correspondingauthor{Thomas Wevers}
\email{twevers@stsci.edu}
\author[0000-0002-4043-9400]{T. Wevers}
\affiliation{Astrophysics \& Space Institute, Schmidt Sciences, New York, NY 10011, USA}
\affiliation{Space Telescope Science Institute, 3700 San Martin Drive, Baltimore, MD 21218, USA}
\author[0000-0002-5063-0751]{M. Guolo}
\affiliation{Bloomberg Center for Physics and Astronomy, Johns Hopkins University, 3400 N. Charles St., Baltimore, MD 21218, USA}
\author[0000-0002-0743-2645]{S. Lockwood}
\affiliation{Space Telescope Science Institute, 3700 San Martin Drive, Baltimore, MD 21218, USA}
\author{A. Mummery}
\affiliation{Oxford Theoretical Physics, Beecroft Building, Clarendon Laboratory, Parks Road, Oxford, OX1 3PU, United Kingdom}
\author[0000-0003-1386-7861]{D.R. Pasham}
\affiliation{MIT Kavli Institute for Astrophysics and Space Research, 70 Vassar Street, Cambridge, MA 02139, USA}
\author[0000-0003-4054-7978]{R. Arcodia}
\affiliation{MIT Kavli Institute for Astrophysics and Space Research, 70 Vassar Street, Cambridge, MA 02139, USA}
\begin{abstract}
\noindent
X-ray quasi-periodic eruptions (QPEs) are a novel mode of variability in nearby galactic nuclei whose origin remains unknown. Their multi-wavelength properties are poorly constrained, as studies have focused almost entirely on the X-ray band. Here we report on time-resolved, coordinated {\it Hubble Space Telescope} far ultraviolet and {\it XMM-Newton} X-ray observations of the shortest period X-ray QPE source currently known, eRO-QPE2. We detect a bright UV point source ($L_{\rm FUV} \approx {\rm  few} \times 10^{41}$ erg s$^{-1}$) that does not show statistically significant variability between the X-ray eruption and quiescent phases. This emission is unlikely to be powered by a young stellar population in a nuclear stellar cluster. The X-ray-to-UV spectral energy distribution can be described by a compact accretion disk ($R_{\rm out} = 343^{+202}_{-138} \ R_{\rm g}$). Such compact disks are incompatible with typical disks in active galactic nuclei, but form naturally following the tidal disruption of a star. Our results rule out models (for eRO-QPE2) invoking i) a classic AGN accretion disk and ii) no accretion disk at all. For orbiter models, the expected radius derived from the timing properties would naturally lead to disk-orbiter interactions for both quasi-spherical and eccentric trajectories. We infer a black hole mass of log$_{10}(M_{\rm BH}) = 5.9 \pm 0.3$ M$_{\odot}$ and Eddington ratio of 0.13$^{+0.18}_{-0.07}$; in combination with the compact outer radius this is inconsistent with existing disk instability models. After accounting for the quiescent disk emission, we constrain the ratio of X-ray to FUV luminosity of the eruption component to be $L_{\rm X} / L_{\rm FUV} > 16-85$ (depending on the intrinsic extinction).  
\end{abstract}

\section{Introduction} \label{sec:intro}
Quasi-periodic X-ray eruptions (QPEs) are a recent addition to the various modes of rapid variability observed in massive black holes (MBHs) inhabiting galactic nuclei. Their X-ray timing and spectral properties, including quasi-periodic, high-amplitude outbursts and the emergence of an additional hot thermal component during the outburst rise, are distinct among the known variability of active galactic nuclei (AGN; \citealt{Miniutti19, Giustini20, Arcodia21}). 

The nature of QPEs remains the subject of debate, but may relate to accretion disk instabilities \citep{Sniegowska20, Raj21, Pan23}, or to the interaction between the supermassive black hole (or an accretion disk surrounding it) and a stellar-mass companion. The latter class of models comes in many flavors, including repeated partial tidal disruptions of stellar-mass objects \citep{King22}, stable or unstable Roche-lobe overflow \citep{Krolik22, Metzger22, Luquataert23} and star- or BH-disk interactions \citep{Dai2010, Linial23, Franchini23, Tagawa23,Zhou24}. 

In addition to their unique properties, QPE sources appear to exhibit some characteristics that can be linked to tidal disruption events (TDEs). This includes declining long-term--albeit sparsely sampled--lightcurves for some sources \citep{Miniutti23, Arcodia24}, QPE-like flares observed in TDE candidates \citep{Chakraborty21, Quintin23, Bykov24}, and an over-representation of host galaxies with post-starburst characteristics \citep{French16, Wevers22} as well as extended emission line regions \citep{Wevers24a, Wevers24b} in both QPEs and TDEs. These similarities also extend to the morphological properties of the host galaxies \citep{Gilbert24}. The recent detection of QPEs following a spectroscopically confirmed TDE \citep{Nicholl2024} has provided a direct link between (at least some) TDEs and QPEs.

Most existing QPE models have been developed based on the observed X-ray properties of QPEs, with very few constraints available at other wavelengths. Consequently it is challenging to falsify existing models using X-ray data alone, and very few predictions exist for the expected properties of QPEs at other wavelengths. Notable examples include \citet{Linialuvqpe}, who predicted that in the star-disk collision model active X-ray QPE sources should not exhibit UV QPEs, as the parameter space for these to be observable is distinct (in terms of black hole mass and accretion rate). \citet{Vurm24} quantified the emission of this model further by performing radiation transport calculations, providing detailed predictions for the QPE flare spectral energy distribution (SED). These model predictions are consistent with the lack of UV variability in existing datasets obtained by the Optical Monitor (OM) telescope on \xmm (e.g. \citealt{Arcodia21}). However, due to the very large point spread function (PSF) and modest telescope size, these measurements encompass most of the host galaxy light and therefore do not have much constraining power to transient QPE emission. 

In this work we report on deep, time-resolved observations at far ultraviolet (FUV) wavelengths obtained with the {\it Hubble Space Telescope} (\hst). Together with coordinated \xmm X-ray observations, these provide the deepest FUV constraints of an X-ray QPE source to date. A detailed analysis of the (time-resolved) X-ray properties of eRO-QPE2 has been reported in \citet{Arcodia24qpe2} and \citet{Pasham24}, and we focus on novel information that the UV data provides in combination with the X-ray observations in this work. We present the observations and analysis in \S\ref{sec:analysis}. The results are presented in \S\ref{sec:discussion}, where we discuss their implications in detail for a number of QPE model scenarios. We summarize the main results in \S\ref{sec:summary}.

\section{Observations and Analysis}
\label{sec:analysis}
\subsection{Observations}
We obtained coordinated observations of the QPE source with the shortest recurrence time (the time between the peaks of consecutive eruptions) that is currently known, \target, with sky coordinates in decimal degrees (ra, dec) = (38.70300, --44.32569) \citep{Arcodia21}. The host galaxy redshift is $z=0.0175$ \citep{Arcodia21}, corresponding to an approximate physical scale of 360 pc arcsec$^{-1}$. X-ray observations were taken with \xmm, while UV observations were performed with the Space Telescope Imaging Spectrograph (STIS) onboard \hst. 
We used the STIS Far-Ultraviolet Multi-Anode Microchannel Array (FUV-MAMA) detector in combination with the F25QTZ longpass filter (with a pivot wavelength of 1596 \AA), which provides a native time resolution of 125 microseconds and a spatial resolution of 0.0246 arcsec per pixel over a $\sim$25$\times$25 arcsec field of view.

Each visit consists of a 30 ks \xmm observation and 5 contiguous \hst orbits. The observations were taken in two visits separated by 45 days, on 2023 December 20 (visit 1) and 2024 February 4 (visit 2). 
The first orbit of visit 1 was lost due to a failure to acquire the guide star, so this visit consists of 4 orbits with usable \hst data. 

\begin{figure*}
    \centering
    \includegraphics[width=\linewidth]{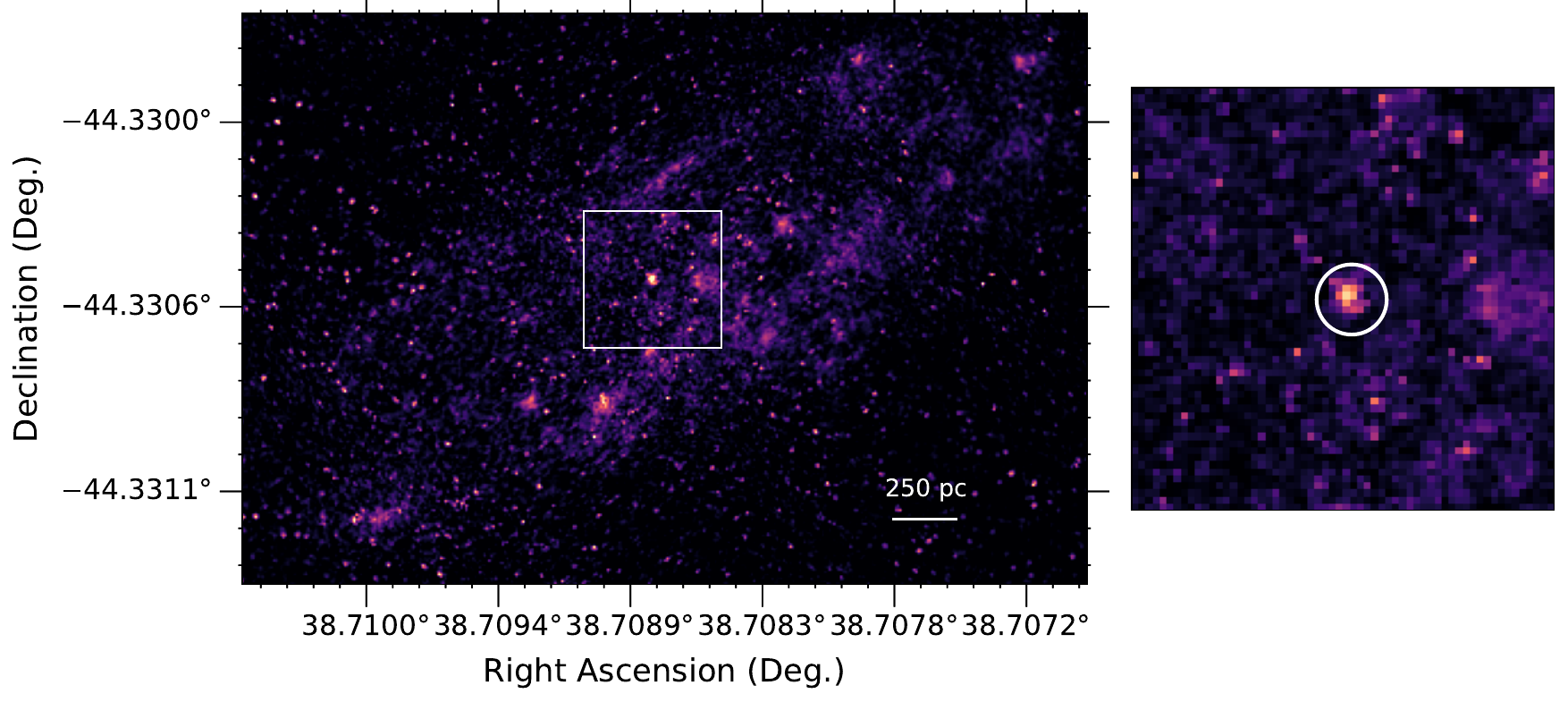}
    \includegraphics[width=0.47\linewidth]{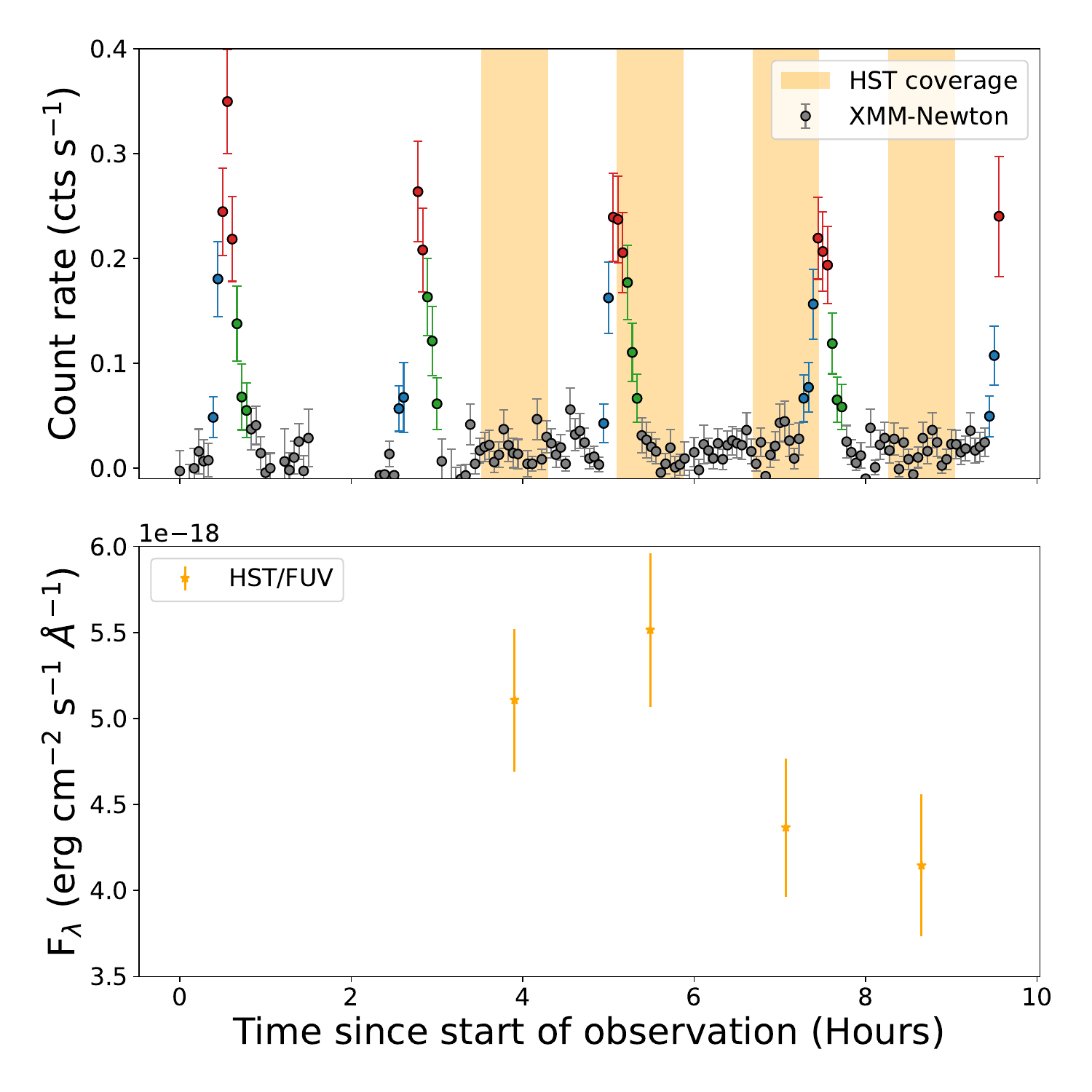}
    \includegraphics[width=0.47\linewidth]{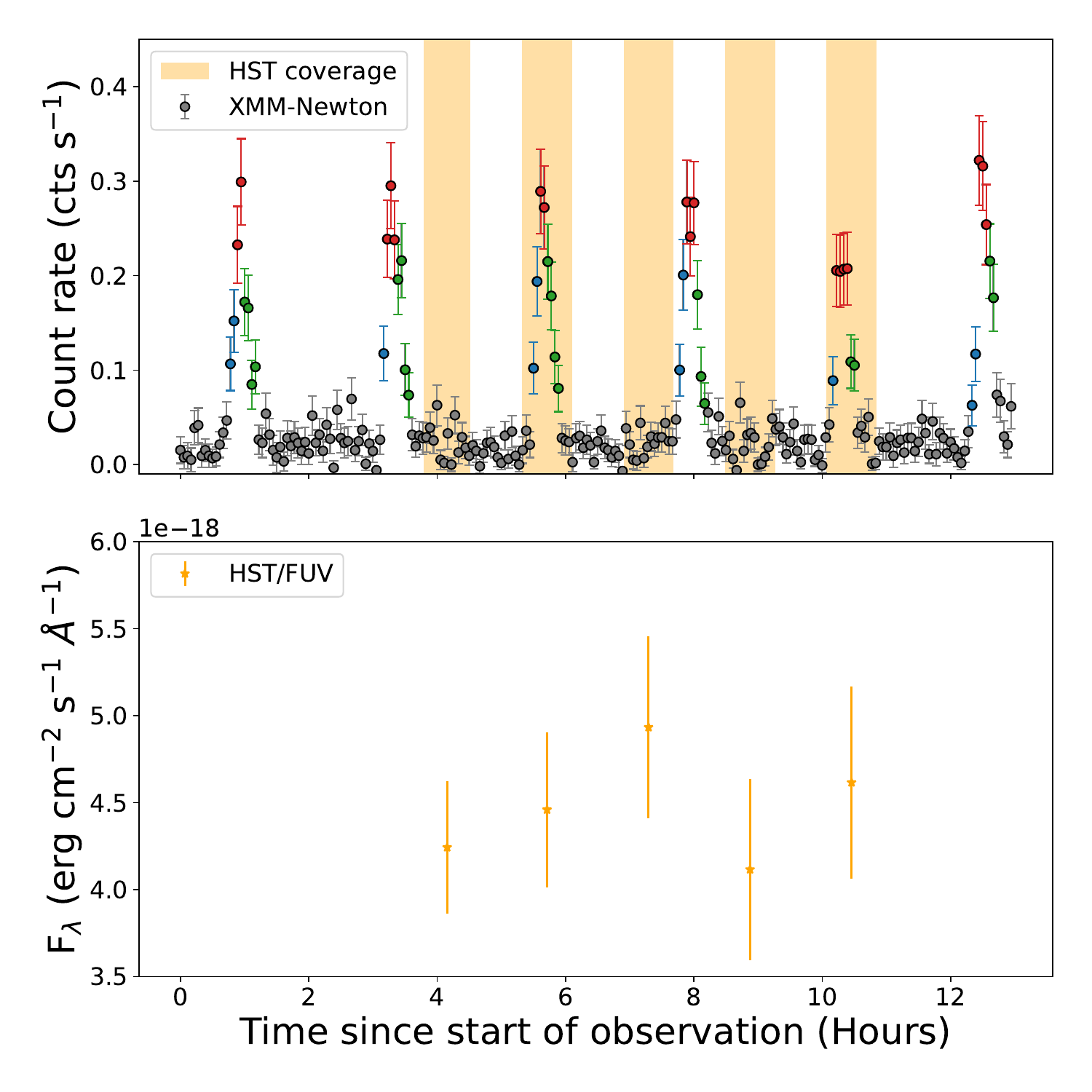}
    \caption{{\bf Image and lightcurves of eRO-QPE2.} The top left panel shows an FUV image of the eRO-QPE2 host galaxy, while the top right panel shows a zoom-in of the white box (0.75 arcsec on a side), indicating the 0.122 arcsec aperture used to perform the photometry of the nucleus. The middle panels show the X-ray lightcurves overlaid with the \hst on-source periods in orange. Grey datapoints indicate quiescence, while blue/red/green point indicate eruption rise/peak/decay, respectively. The bottom panels show the FUV measurements made on the per-orbit stacked images with 1--$\sigma$ uncertainties. No statistically significant variability is evident, regardless of whether the X-ray QPE is active or not.}
    \label{fig:fig1}
\end{figure*}

A detailed description of the data reduction of both the X-ray and UV data can be found in the Supplementary Materials. 
We show an FUV 5-orbit stacked image of the entire host galaxy of \target in the top panel of Figure \ref{fig:fig1}. Star-forming regions are seen throughout the galaxy, consistent with emission line maps obtained with MUSE \citep{Wevers24b}.

The middle panels of Figure \ref{fig:fig1} show the \xmm observations (blue) and the on-source periods of the \hst observations (orange shaded regions). During the first visit, roughly half of two eruptions are covered by \hst, while the remaining data covers QPE quiescence. In the second visit, two full eruptions are covered with \hst while the 3 remaining orbits cover the quiescent phase.

Note that due to effects of detector dark glow (described in detail in \S\ref{sec:glow} of the Supplementary Materials), we defer exploring the full time resolution of the observations to the future and use only the orbit-averaged data in this work.

\subsection{Photometry}
We perform photometry centered on the galaxy nucleus with an aperture of 0.122 arcsec, and use the appropriate aperture correction of 0.659 based on a point source PSF \citep{Proffitt2003} to calculate the object brightness in each image. The background contribution is measured from an annulus with inner and outer radii of 30 and 80 pixels (0.74 -- 2 arcsec), chosen to exclude the majority of the encircled energy area of a point source, but including the diffuse galaxy emission within which the nucleus is embedded. We use the median value within this annulus as the local background estimate.

Using the stacked images of the two visits, we detect an FUV point source in the galaxy nucleus with a brightness of $F_{\rm FUV}$ = 4.8$\pm$0.6 10$^{-18}$ erg cm$^{-2}$ s$^{-1}$ \AA$^{-1}$\ (m = 24.90$\pm$0.13 AB mag) in visit 1 and $F_{\rm FUV}$ = 4.5$\pm$0.3 10$^{-18}$ erg cm$^{-2}$ s$^{-1}$ \AA$^{-1}$\ (m = 24.96$\pm$0.06 AB mag) in visit 2. These measurements are consistent within the uncertainties. This point source is also detected in the per-orbit stacked images (see bottom panels of Fig. \ref{fig:fig1}). 

No significant change in brightness is found between the orbits that cover the QPE phase relative to those that cover the quiescent phase. 
Consequently, we obtain the most stringent constraints on the UV variability of X-ray QPEs between eruption and quiescence down to a level of 1.8$\times$10$^{-18}$ (0.9$\times$10$^{-18}$) erg cm$^{-2}$ s$^{-1}$ \AA$^{-1}$\ (at the 3 $\sigma$ level for visit 1 and 2, respectively).  
For reference, this is a factor of $\sim$100 deeper than the constraints provided by the \xmm's Optical Monitor (OM) observations reported in \citet{Arcodia21}. 

\subsection{Extinction correction from the Balmer decrement}
To estimate the intrinsic brightness of the detected point source, an extinction correction is required. This is especially important in the FUV regime explored here. In addition to the Galactic extinction estimated from \citet{Schlafly2011} of E(B--V) = 0.015, two independent measures of the extinction are available, through optical spectroscopy and X-ray spectroscopy (assuming a gas-to-dust ratio), although we caution that both are subject to significant uncertainty.

For the former, we use optical integral field spectroscopy with the Multi Unit Spectroscopic Explorer (MUSE) to constrain the Balmer decrement measured towards the nuclear region of the galaxy. A detailed analysis of these data is presented in \citet{Wevers24b}, and we follow the same procedures to measure the emission line fluxes with the Penalized Pixel Fitting routine \citep[\texttt{pPXF},][]{Cappellari2017}. We measure an average Balmer decrement of $H_{\alpha} /  H_{\beta}$ = 4.07$\pm$0.07 by using a range of apertures radii from 1 pixel to the typical PSF full width at half maximum (FWHM) of the observations (0.7\arcsec) in 0.1 arcsec increments.
By assuming the \citet{Calzetti2000} attenuation law and case B recombination, this measurement can be converted into a color excess E(B--V)
\begin{equation}
    E(B-V) \approx 1.99\ \rm log_{10} \left ( \frac{H_{\alpha}/H_{\beta}}{2.86} \right )
\end{equation}

Under these assumptions we infer E(B--V) = 0.31$\pm$0.04\footnote{This value is consistent with an independent estimate of the optical extinction obtained with \texttt{Bagpipes} \citep{Wevers24b}.}, and we derive an FUV luminosity for the nuclear point source of $L_{\rm FUV} = 1.0^{+0.4}_{-0.3}  \times 10^{41}$ erg s$^{-1}$. Here we have assumed a flat $\lambda$-CDM cosmology with H0 = 73 km s$^{-1}$ Mpc$^{-1}$ \citep{shoes}.

We will discuss the X-ray extinction estimate in \S\ref{sec:sedmodel}, as it is model-dependent and assumes that the UV and X-ray emission originates in an accretion disk. \\

\subsection{A nuclear stellar cluster origin for the FUV emission is very unlikely}
\label{sec:nsc}
A nuclear starburst can in principle create an UV-bright, centrally concentrated population of (young) stars. To constrain whether a nuclear stellar cluster (NSC) can explain the observations, we attempt to reproduce the FUV luminosity using simple stellar population models from \citet{Maraston05}. We conservatively use the smallest extinction correction of E(B--V) = 0.31; if the true correction is closer to the full SED fitting estimate of E(B--V) = 0.5, that would exacerbate the problems of this interpretation. 

For a stellar population age of 100 (10) Myrs, reproducing the FUV luminosity ($L_{\rm FUV}$ = 1$\times$10$^{41}$ erg s$^{-1}$) would require a stellar mass of young stars of log$_{10}$(M$_{\rm young\, stars}$) $\sim$ 6.8 (5.4) M$_{\odot}$. The mass fraction and total mass of recently formed stars was estimated using the Bagpipes full spectrum fitting code in \citet{Wevers24b}, who found a burst fraction of $\sim$10\% and a total stellar mass of 10$^{8.1}$ M$_{\odot}$ contained within a 250 pc aperture. An important caveat is that the spatial resolution of the MUSE spectra is $\sim$0.7\arcsec\ (250 pc physical scale) i.e. much larger than the HST data used here. Assuming this mass fraction would imply a total NSC mass of log$_{10}$(M$_{NSC}$) $\sim$ 7.8 (6.4) M$_{\odot}$, similar to the total mass formed in a putative recent starburst. This would suggest that nearly all the mass must be contained in the NSC, but the bulge/nuclear component is clearly resolved in DESI Legacy Survey continuum imaging \citep{desi}. The total NSC mass would also make it a very massive system \citep{Neumayer20}, while the required mass of young stars would make this a system with one of the highest masses of young stars in NSCs (e.g. \citealt{Seth06}). 

Moreover, the SFR surface density would have to be extremely high. Assuming that the point source subtends $<$0.1 arcsec\footnote{The 50\% encircled energy radius of the F25QTZ filter is 0.073 arcsec \citep{Proffitt2003}.} (which corresponds to 36 pc in physical scale), and that the young stellar population formed on a timescale similar to its age, the SFR surface density ($\Sigma_{\rm SFR}$) would be of $\mathcal{O}$(10 M$_{\odot}$ yr$^{-1}$ kpc$^{-2}$).
However, the typical $\Sigma_{\rm SFR}$ of star-forming regions in the local universe is $\mathcal{O}(10^{-2} \, \mathrm{M_{\odot} \, yr^{-1} \, kpc^{-2}})$. An analysis of spaxels in the SDSS MaNGA survey \citep{law2022} shows that $\sim 0.5\%$ of these spaxels have $\Sigma_{\rm SFR} \geq 0.1 \, \mathrm{M_{\odot} \, yr^{-1} \, kpc^{-2}}$, and none exceed $\sim 1 \, \mathrm{M_{\odot} \, yr^{-1} \, kpc^{-2}}$. Attributing the inferred FUV luminosity to a NSC in \target\ would require its NSC to have a $\Sigma_{\rm SFR}$ more than an order of magnitude higher than the most extreme star-forming spaxels observed in the MaNGA survey.

In Figure \ref{fig:disk_sed}, we present the maximum FUV luminosity predicted by simple stellar population models at ages of 10 Myr and 100 Myr, corresponding to the highest star formation rate surface density ($\Sigma_{\rm SFR} \sim 1 \, \mathrm{M_{\odot} \, yr^{-1} \, kpc^{-2}}$) observed in the MaNGA survey \citep{law2022}. These models indicate that even a NSC with the highest $\Sigma_{\rm SFR}$ can account for less than 10\% of the FUV luminosity. Consequently, any NSC contribution would constitute a subdominant fraction to the observed FUV emission. Our modeling results are hence insensitive to the specific properties of a potential NSC.

\subsection{Spectral energy distribution modeling}
\label{sec:sedmodel}
Given that the nuclear FUV point source is unlikely to be of stellar origin (\S\ref{sec:nsc}), it may be related to the quiescent (in-between eruption) X-ray emission. The X-ray quiescent emission of QPEs is usually associated with, and well described by, the inner emission of a standard radiatively efficient accretion disk \citep{Miniutti19, Giustini20, Arcodia21}, similar to that observed in X-ray bright TDEs (e.g. \citealt{Mummery2023, Guolo2024}). In this scenario the UV emission would be associated with the cooler mid-to-outer parts of the disk. The joint fitting of the X-ray and UV/optical emission can then be used to constrain the extent of the disk \citep[e.g.,][]{Mummery2020,Nicholl2024,mummery2024fitted, Guolo2024disksed}. 

For detailed modeling under the assumption of an accretion disk origin of the emission, we extract \xmm spectra of various phases of the eruptions. As eruptions were only covered by \hst orbits in visit 2, we restrict all analysis below to data from that observation. Good time intervals are defined to extract the quiescent emission as well as the rise, peak and decay of each of the six eruptions (see the color-coding in Fig. \ref{fig:fig1} middle panel). 

We quote the median value and parameter range containing 68\% of the posterior distribution as the measured value and its uncertainty, unless indicated otherwise.

\subsubsection{Quiescent phase}
We simultaneously and self-consistently fit the FUV photometry and the quiescent X-ray spectrum with the recently developed disk model 
\texttt{diskSED} \citep{Guolo2024disksed}. The \texttt{diskSED} implementation is a standard \citep{Shakura1973} thin disk model with a null-stress boundary condition, that includes the effects of radiative transfer in the atmosphere of the disk from electron scattering and metal opacity effects via a temperature-dependent color correction factor \citep[$f_{\rm c}$,][]{Shimura1995,Hubeny2001}. The model is suited for
broadband (X-ray spectra + UV/optical/NIR photometry) SED fitting as, in addition to standard inner disk parameters, the outer edge/radius of the disk 
(\Rout) can also be marginalized over. The three free parameters of the model are 
the peak physical temperature of the disk (\Tp),  \Rins ($=R_{\rm in}\sqrt{ {\rm cos} \, i}$) -- where 
\Rin is the inner radius of the disk and $i$ is the inclination of the disk with respect to 
the observer -- and the dimensionless size of the disk (\Rratio). 
Allowing for intrinsic extinction introduces an additional free parameter, the color excess E(B--V). 
Full details of the fitting methodology are provided in \S\ref{sec:fit}.

\begin{figure*}
    \includegraphics[width=\textwidth]{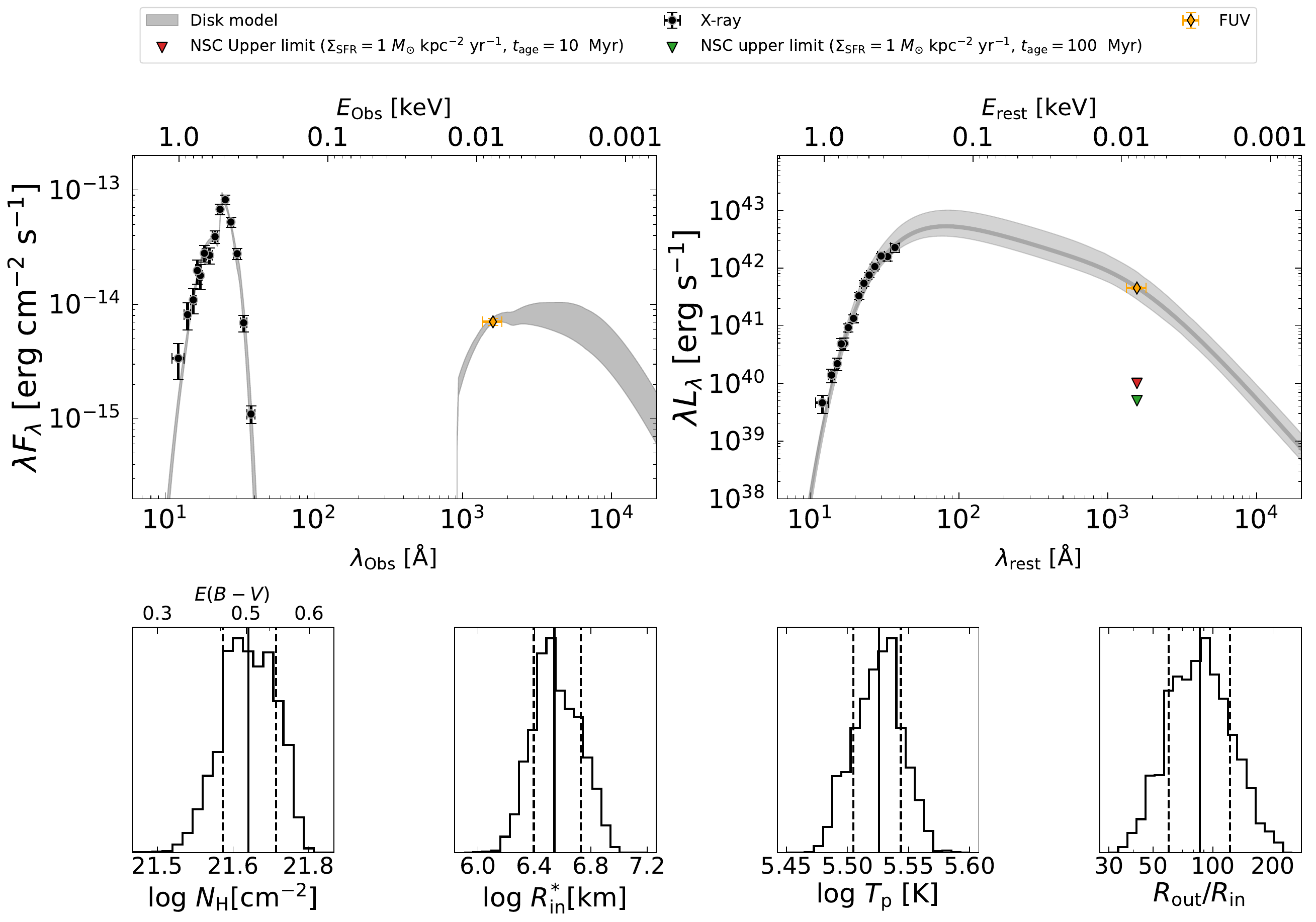}
    \caption{{\bf Quiescent SED modeling results.} The observed (top left) and extinction-corrected, rest-frame (top right) SEDs are shown (black: X-ray data, orange: FUV data). The shaded regions show the best-fit model with the grey band indicating region containing 68\% of the model posteriors. Also shown are upper limits for young stars (green: 10 Myr, red: 100 Myr) in a NSC, using simple stellar populations models (see \S\ref{sec:nsc} for details), assuming the maximum star formation rate surface density in SDSS MaNGA survey. In both panels the data is unfolded to the median of the posterior. Marginalized posterior distributions are shown for all free parameters in the bottom row, where solid lines indicate the median and dashed lines the 68\% credible interval. }
\label{fig:disk_sed}
\end{figure*}

Our model for the joint fit of FUV photometry and X-ray spectra has 4 free
parameters, for which we assume uniform (or log-uniform) priors. In \texttt{XSPEC}, this model can be generated as \texttt{phabs$\times$redden$\times$zashift(phabs$\times$reddenSF$\times$diskSED)}. The fitting described here is performed together with the procedure described in \S\ref{sec:erup_SED} for the eruption phases for self-consistency, such that the derived intrinsic $N_{\rm H}$ can be tied between the phases.

The results of the nested sampling fit are shown in Fig.~\ref{fig:disk_sed}. The bottom panel shows the 1D projection of the parameter posterior distributions. The full posteriors of all parameters are shown in Fig. \ref{fig:qpe2_post}. In the upper left panel of Fig.~\ref{fig:disk_sed}, we show the observed flux model (without extinction/absorption corrections) overlaid on the observed FUV photometry and the unfolded X-ray spectrum. The right panel shows the intrinsic luminosities (with both Galactic and intrinsic absorption/attenuation corrections), with the data points unfolded to the median values of the parameter posteriors. The red and green triangles show upper limits on the expected emission of young stellar population in a putative NSC for very high SFRs (see \S\ref{sec:nsc} for details).

\subsubsection{Eruption phase}\label{sec:erup_SED}
We also obtain constraints on the SED of the eruptions during rise, peak and decay phases through a similar modeling approach. To this end we add a thermal component to the model used to describe the quiescent emission. This additional component describes the hotter/excess emission during the eruptions well (\texttt{bbody} in \texttt{XSPEC}, e.g. \citealt{Miniutti19}). 

\begin{figure*}
    \includegraphics[width=\textwidth]{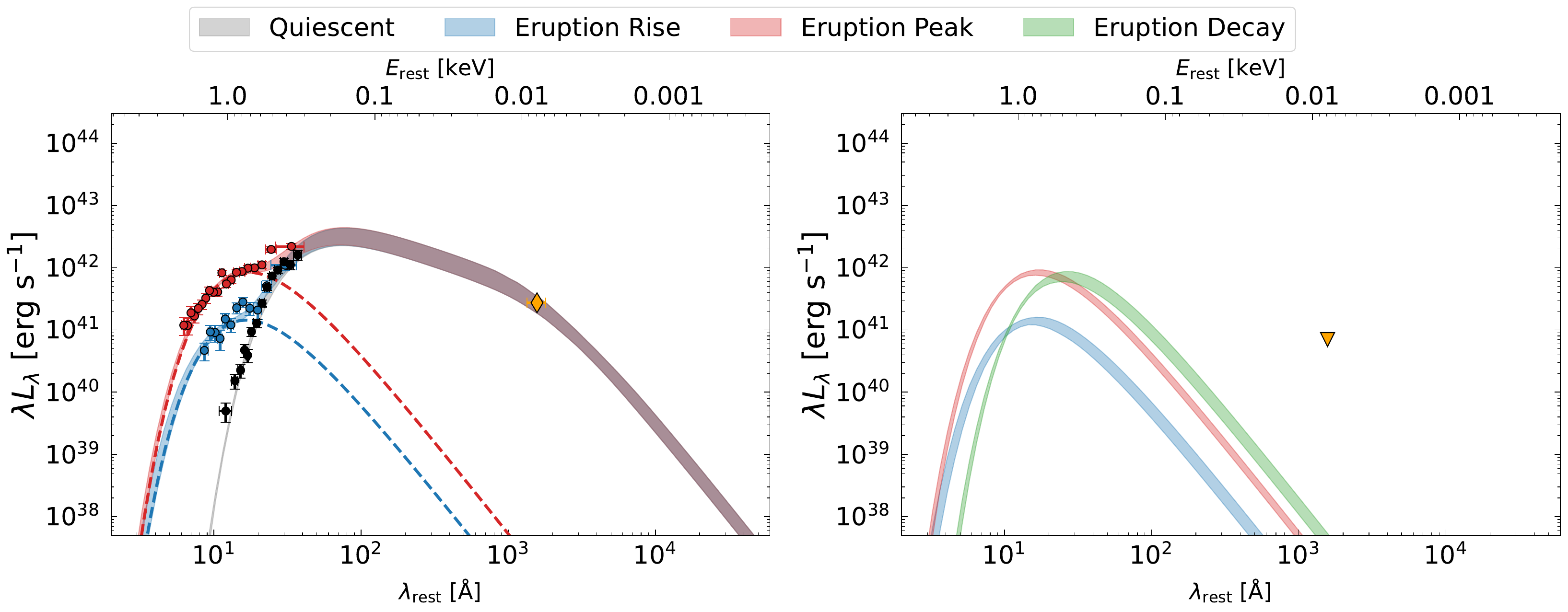}\\
    \includegraphics[width=\textwidth]{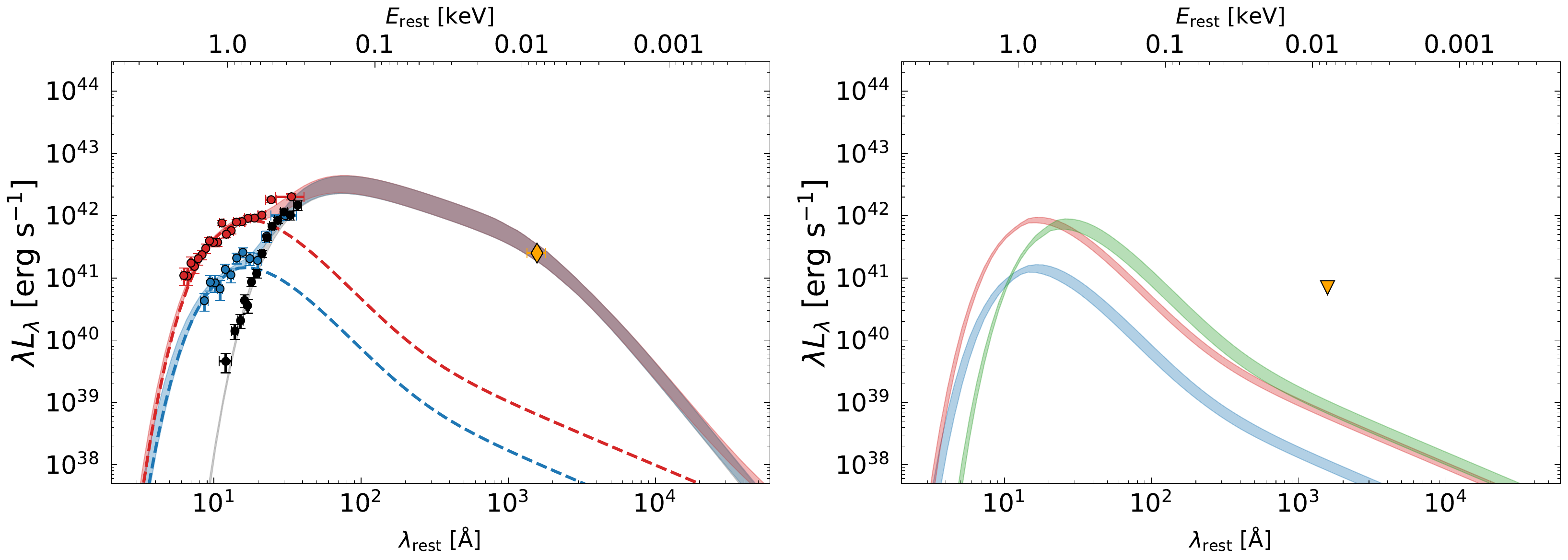}
    \caption{{\bf Constraints on the eruption SED.} The top left panel shows the X-ray spectra in the quiescent (black/gray), rise (blue) and peak (red) phases. The decay phase is omitted for clarity. Shaded regions illustrate the 68\% CI of the posterior distributions. Dashed lines show the median values of the posteriors for the thermal/\texttt{bbody} component. The top right panel illustrates the 68\% confidence contours of the posteriors of the thermal component only; a 3--$\sigma$ upper limit is shown in the FUV (i.e. this assumes that the FUV emission is produced entirely by the quiescent emission). The bottom panels are identical to the top panels but show the more detailed calculations from \citet{Vurm24} (see text for details).}
\label{fig:eruption_sed}
\end{figure*}

Following a similar fitting procedure, we show the SED of the quiescent (black/gray), rise (blue) and peak (red) phases in Figure \ref{fig:eruption_sed}. The right hand panel shows the eruption component only, which then leads to an upper limit in the FUV. These are the first multi-wavelength constraints on the SED of the variable spectral component that is responsible for the QPEs. 

In addition to a simple single-temperature thermal component, we also include results from a spherically symmetric Monte Carlo radiation transport simulation which follows the production of photons behind the radiation-mediated shock, Comptonization by hot electrons, and the eventual escape of the radiation through the expanding debris in the disk-star interaction model \citep{Vurm24}. The resulting SED from the \citet{Vurm24} simulations in the wavelength range of interest (X-ray and UV) can be written to first order as:

\begin{equation}
     F_{\nu} (\nu, T_{obs}) \approx B_{\nu} (\nu, T_{obs}) + 0.1 B_{\nu} (\nu_{peak}, T_{obs}) \mathcal{H} (\nu_{peak}- \nu )
\end{equation}

\noindent where $B_{\nu} (\nu, T_{obs})$ is a Planck function with observed temperature  $T_{obs}$, 
$\nu_{peak}$ is the peak frequency of Planck function, and $\mathcal{H}$ is the Heaviside function.
We implement this analytical form of the SED into a \texttt{pyXspec} model, and add it instead of \texttt{bbody} to fit the additional emission of the eruptions. The results are shown bottom panel of Fig. \ref{fig:eruption_sed}.

Both of these models remain consistent with our FUV constraints. We note that a high extinction correction is required in this system; for a similar system without extinction our FUV constraints would have been a factor of $\sim$30 deeper. 
After accounting for the quiescent disk emission, the peak X-ray luminosity of the eruption component is log$_{10}$($L_{\rm peak, X}$) = 42.15$\pm$0.10 erg s$^{-1}$, and the upper limit on the FUV luminosity is $L_{\rm max, FUV} < $ 9$\times$10$^{40}$ erg s$^{-1}$ (3 $\sigma$, assuming E(B--V) = 0.5; if E(B--V) = 0.3 as inferred from the Balmer decrement, this upper limit would decrease by a factor of $\approx$5 to $L_{\rm max, FUV} < $ 1.7$\times$10$^{40}$ erg s$^{-1}$ ). The ratio of luminosities which any theoretical model predicts for the eruption component should therefore be $L_{\rm peak, X}$ / $L_{\rm max, FUV} > 16 $ (85 for E(B--V) = 0.3). 

\section{Results and Discussion}
\label{sec:discussion}
We have detected a bright FUV point source in the nucleus of the \target host galaxy, and we argued that this is very unlikely to originate from a nuclear stellar cluster (\S\ref{sec:nsc}). Instead, a compact accretion disk can self-consistently explain the UV to X-ray SED. From the SED modeling, we infer a neutral hydrogen column density of ${\rm log}(N_{\rm H}) = 21.60_{-0.05}^{+0.10} $ cm$^{-2}$, or E(B--V) = 0.50$^{+0.05}_{-0.10}$ assuming a standard gas/dust ratio of 100. The inferred FUV luminosity is then $L_{\rm FUV} = 5 \pm 3 \times 10^{41}$ erg s$^{-1}$. Note that the inferred E(B--V) from the SED modeling (E(B--V) $= 0.50^{+0.05}_{-0.10}$) is high compared to the value obtained from the Balmer decrement (E(B--V) $= 0.31 \pm 0.04$). This tension can be resolved if the host galaxy has a gas/dust ratio that differs from the standard assumption of 100 \citep{Guver09}. Specifically, if there is more gas relative to the dust, that would decrease the E(B--V) inference from the SED modeling. The host galaxy of eRO-QPE2 is known to be gas-rich (based on the presence of extended ionized gas emission line regions in IFU spectroscopy, \citealt{Wevers24b}), which is consistent with our results.

Black hole mass (\Mbh) estimates can be obtained from the \texttt{diskSED} \Rins parameter — under assumptions for inclination ($i$) and dimensionless spin ($a$) — by associating \Rin with the innermost stable circular orbit (ISCO), using the following expression \citep{Guolo2024disksed}:

\begin{equation}\label{eq:M_BH}
    M_{\rm BH} = \frac{R_{\rm in}^* c^2 }{\gamma(a) G \sqrt{{\rm cos}\, i}}.
\end{equation}

\noindent where $\gamma(a)$ is the ISCO location in gravitational radii, which is a function of the spin \citep[see e.g.,][]{Bardeen72}, such that $\gamma(0) = 6$ and $\gamma(1) = 1$. 

We assume a flat probability distribution of prograde spins in the $0 \leq a \leq 0.99$ range as well as the full range of values ($-0.99 \leq a \leq 0.99$), and a flat probability distribution for ${\rm cos} \,i$, with inclinations in the range $0^\circ \leq i \leq 80^\circ$\footnote{The reason to limit it to $80^\circ$, is that the Newtonian photon propagation approximation of \texttt{diskSED} likely breaks down in more edge-on cases.}. This results in a \Mbh estimate of log(\Mbh) = $5.9\pm 0.3$ \msun (see Supp. Mat. \S\ref{sec:mbh} for a comparison to alternative estimates). The resulting probability distribution of \Mbh including the uncertainties driven by the (unknown) $a$ and $i$ are shown in Fig.~\ref{fig:dynamics} (bottom left panel). 

The ratio of the outer to inner disk radius is \Rratio$= 86^{+36}_{-25}$, corresponding to an outer disk radius of $R_{\rm out} = 343^{+202}_{-138} R_g$, for the assumed flat probability distribution of spins. This is of the same order of magnitude as the two other QPE systems for which this estimate is available (AT2019qiz, \citealt{Nicholl2024} and GSN069, \citealt{Guolo2025}.). The peak physical temperature of the disk is {\rm  log(\Tp)$= 5.50 \pm 0.05$ K.  
The multi-wavelength modeling provides a more accurate estimate of the bolometric disk luminosity compared to X-ray-only estimates, yielding $L_{\rm Bol} = 1.3^{+1.0}_{-0.5} \times 10^{43}$ \ergs. This translates into an Eddington ratio of the quiescent accretion disk emission $\lambda_{\rm Edd} = 0.13^{+0.18}_{-0.07}$. The quoted range includes statistical uncertainties as well as uncertainties introduced by the derived black hole mass, which already account for the effects of the assumed spin and inclination distributions. These results are insensitive to potential systematic uncertainties of (for example) a small (1--10 per cent) contribution of an underlying NSC to the FUV emission.

We now discuss the implications of these results in the context of theoretical models for QPEs.

\subsection{On the origin of a compact accretion disk}
Our modeling demonstrates that the bright FUV point source that is detected is consistent with an extrapolation of the X-ray spectrum in the assumption of a compact accretion disk origin. By modeling the X-ray and UV SED, we infer that this accretion disk is compact, $R_{\rm out} = 343^{+202}_{-138} R_g$. This is similar to the expected size for viscously spreading disks following the tidal disruption of a star \citep{Cannizzo1990, Mummery2020}, but grossly inconsistent with the typical sizes of accretion disks in AGNs, which are persistent and
source their material from large radii ($\sim 10^5$ R$_g$). Consequently we can rule out all models associated with standard AGN accretion disks as the origin of the X-ray QPEs in eRO-QPE2. These results are consistent with the absence of broad emission lines in optical spectroscopy \citep{Wevers22, Wevers24b} as well as an X-ray corona and an IR bright torus-like structure, indicating that no mature broad line region is present \citep{Miniutti19}. Models that do not invoke an accretion disk at all are also strongly disfavored to explain the QPEs in \target.

Both the size and luminosity are typical of accretion disks formed in the aftermath of TDEs \citep{vanvelzen2019,Mummery2024}, although the time since any TDE remains unconstrained and the typical cooling observed in TDE disks at late times is not seen in eRO-QPE2 \citep{Arcodia2024qpe2, Pasham24}. The latter finding may indicate that the disk mass is replenished in some way (e.g. stellar ablation, \citealt{Linial23, PYao24}), or that the disk has an extremely long `viscous' time-scale, as compared to most known TDE disks.

\begin{figure*}
    \centering
    \includegraphics[width=0.9\textwidth]{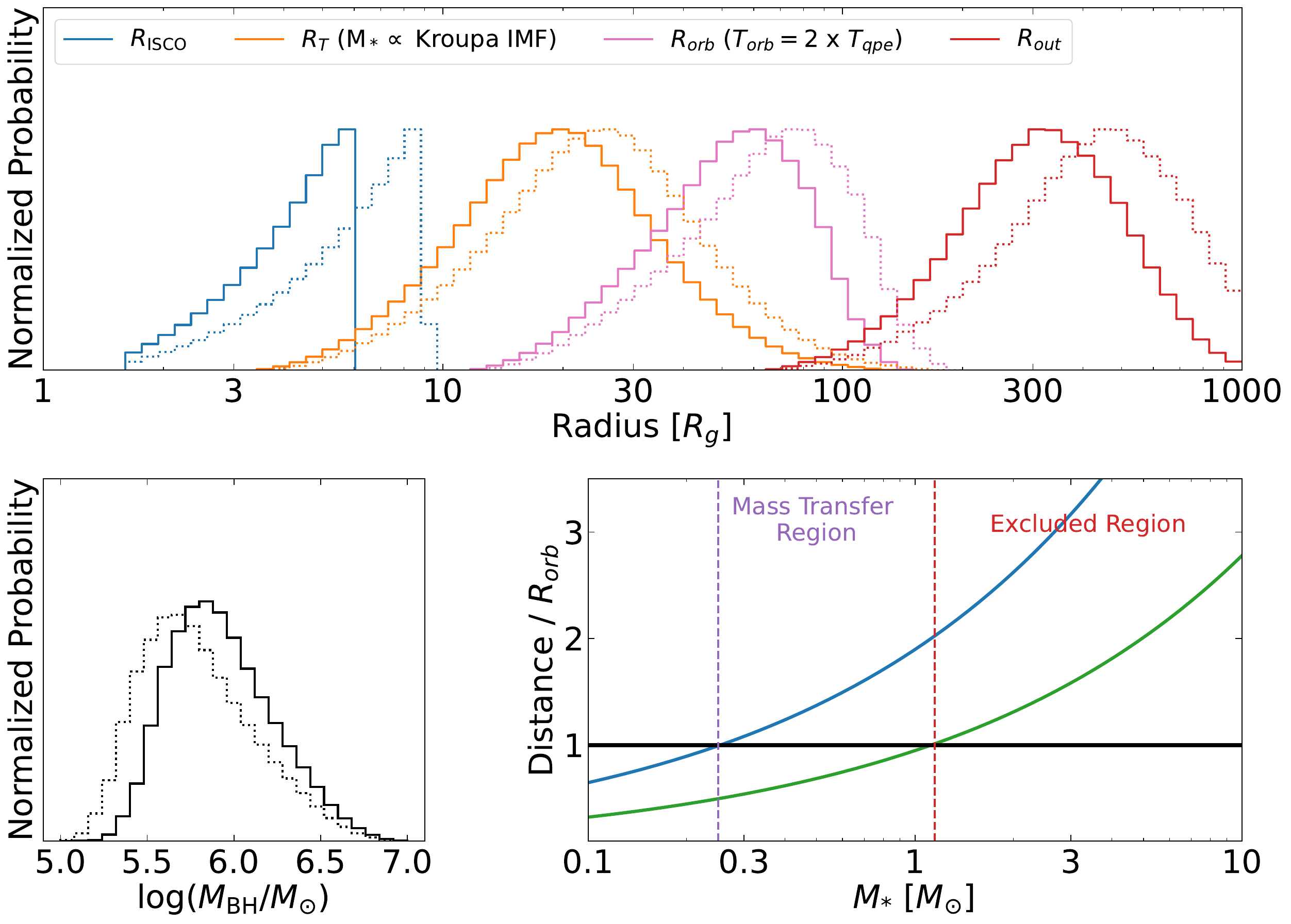}
    \caption{{\bf Relevant length scales and parameter inferences.} The top panel shows probability density functions of the various length scales in the system, including the black hole ISCO radius (blue), the stellar tidal radius (orange), the orbiter radius (green) and the disk outer radius (red). Dotted lines show the posteriors for the full spin range, while the solid lines show the results for positive spin. The bottom left panel shows the posterior distribution of the inferred black hole mass. The bottom right panel illustrates that the (MS) stellar mass cannot exceed 1 M$_{\odot}$, as this would lead to a full tidal disruption (see text for more details).}
\label{fig:dynamics}
\end{figure*}

\subsection{Constraints on UV QPE variability}
Our time-resolved observations provide the most stringent constraints to date on the UV variability during X-ray eruptions. The limiting factor is the presence of a bright FUV point source coincident with the galaxy nucleus with a luminosity of $L_{\rm FUV} = {\rm few} \times 10^{41}$ erg s$^{-1}$; we are insensitive to lower amplitude variability.

Our observations also provide the first multi-wavelength constraints on the SED of the QPE flares (Fig \ref{fig:eruption_sed}). Both a simple blackbody extrapolation of the X-ray spectrum into the UV (top panels) as well as a more detailed calculation of the UV emission due to disk-orbiter interactions (bottom panels) are consistent with our FUV constraints.

To render potential UV counterparts to the X-ray QPEs detectable with HST for an eRO-QPE2-like system (for disk-orbiter interaction models), the UV luminosity of the accretion disk needs to be suppressed by a factor of $\gtrsim$100. This may be feasible for a low black hole mass, low accretion rate systems \citep{Linialuvqpe} if HST-like sensitivities can be achieved (e.g. by the Ultraviolet Explorer mission, \citealt{uvex}). 

\subsection{Constraints on QPE orbiter models}
By assuming the SMBH mass obtained from the SED modeling, we can translate various relevant length scales of the system to gravitational radii. These are shown as probability density functions in the top panel of Figure \ref{fig:dynamics}. In blue we show the innermost stable circular orbit (ISCO) radius, which is the distribution of $R_{\rm ISCO}/R_g$ for the assumed flat spin distribution. In orange, we show the stellar tidal disruption radius ($R_{T} \approx R_{\star} (M_{\rm BH}/M_{\star})^{1/3}$) for a main sequence star (i.e, $R_{\star} \propto M_{\star}^{4/5}$) with masses distributed assuming a \citet{Kroupa2001} stellar mass function. The orbital radius (green) is derived from the QPE recurrence time, assuming a quasi-circular orbit and the \Mbh distribution. In red we show the disk outer radius, inferred from the SED modeling. The parameters for positive spin values are shown as solid lines, while the full spin range is shown as dotted lines. The width of these distributions reflects the parameter uncertainties, both those measured from the data (e.g., \Mbh and $T_{QPE}$) and those assumed {\it ad hoc} (e.g., $a$ and $i$).

A fundamental requirement for disk-orbiter interaction models to remain viable is a configuration where the orbit crosses the accretion disk, i.e. the orbital radius must be smaller than the disk outer radius. 
It is immediately evident that in the assumption of quasi-circular configurations, this is the case for eRO-QPE2 and hence such models remain compatible with our FUV observations.

\subsubsection{Nature and size of the orbiter in quasi-spherical configuration}
The results of our SED modeling provide quantitative constraints on both the SMBH and the accretion disk parameters. This allows us to quantitatively consider the range of allowed stellar parameters, and their implications for various models. 

These constraints are visualized in the bottom right panel of Fig. \ref{fig:dynamics}, where we assume a main-sequence (MS) star companion and a quasi-circular orbit (i.e. T$_{\rm orb}$ = 2$\times$T$_{\rm QPE}$), such that the orbital radius $R_{\rm orb}$ relates to the QPE recurrence time as: $T_{\rm QPE} = \pi \sqrt{R_{\rm orb}^3/G M_{BH}}$. We plot the stellar tidal radius as a function of stellar mass (assuming the mass-radius relation of \citealt{Kippenhahn1991}) and distance from the black hole (normalized by the orbital radius) in green. Note that this panel is black hole mass independent. The star cannot cross the tidal radius without being completely destroyed, implying that the orbiter cannot be more massive than $\approx$1 M$_{\odot}$. For less massive stars, those that orbit within 1--2 R$_{T}$ will be susceptible to overflowing their Roche lobes, and as can be seen from the top panel of Fig. \ref{fig:dynamics}, these orbits are also liable to disk crossings. For low mass stars below $\sim$0.25 M$_{\odot}$, the only viable model is the disk-star collision model as such stars will not exhibit Roche-lobe overflow. 

An independent estimate of the mass (and nature) of the companion star would therefore allow us to discriminate between QPE models in the future. Meaningful lower limits may be obtained with continued monitoring observations. For example, in the $\sim$4 years since its discovery eRO-QPE2 has completed $\sim$15\,000 QPE cycles. Assuming that a stellar-like orbiter loses mass through ablation when crossing the disk \citep{Linial24, PYao24} can lead to a lower limit on the object mass. \citet{PYao24} show that for typical parameters, the expected mass-loss per collision is $\mathcal{O}$(10$^{-5}$-- 10$^{-4}$ M$_{\odot}$), although there is a large spread in plausible values depending on the system parameters. With this baseline assumption, in 10$^4$ cycles the star would have lost 0.1--1 M$_{\odot}$ of material through ablation. The fact that QPEs are on-going then leads to a lower limit of $\sim$0.1 M$_{\odot}$ for the original mass of the orbiter, while ablation rates in excess of 10$^{-4}$ M$_{\odot}$ per encounter can be ruled out for eRO-QPE2 (as a star more massive than 1M$_{\odot}$ would have entered its full disruption radius at the observed recurrence time). Note, however, that in reality the star will follow a more complex evolution in response to mass being stripped, potentially changing this picture considerably.

\subsubsection{Accretion disk instabilities}
Several flavors of accretion disk instability models have been proposed to reproduce the timing and spectral properties of QPEs \citep{Cannizzo1993, Sniegowska20, Raj21, Pan23, Kaur24}. In this context the Eddington ratio, black hole mass and disk outer radius of the accretion disk are quantities of interest, as such instabilities typically occur in relatively narrow ranges of this parameter space. We have constrained the Eddington ratio of the accretion flow to be $\lambda_{\rm Edd} = 0.13^{+0.18}_{-0.07}$. At this Eddington ratio the parameter space for thermal/viscous instability is very small for a typical thin disk if magnetic fields are present \citep{Begelman07, Kaur24}. Note that more generally, models invoking accretion rate variations cannot readily explain the temperature hysteresis observed during the eruptions in eRO-QPE2 \citep{Arcodia2024qpe2}.

Disk tearing may occur in strongly warped disks \citep{Nixon12}, although it is unclear which mechanism would operate to induce strong warps given the relative stability of the X-ray emission over 4 years. 

The presence and recurrence time of disk pressure instabilities is set by radiation pressure or magnetic field strength \citep{Kaur2023}, as well as the accretion rate. As already noted in \citet{Arcodia21}, reproducing the timescales involved for eRO-QPE2 would require extreme values of disk viscosity. Using equation 34 from \citet{Grzdzielski2017} to interpolate the recurrence time for radiation pressure instabilities, and assuming an eruption amplitude relative to quiescence of $\gtrsim 10$ leads to an expected recurrence time of $\sim 1500$\,days, inconsistent with the observations.

For magnetic instabilities, we use Eq. 14 from \citet{Kaur2023} to estimate the product of the dimensionless magnetic pressure scaling parameter $p_0$ and the $\alpha$ parametrization of the viscosity $p_0^{56/37} \ \alpha^{30/37} \approx 3.5 \times 10^{4}$ for the observed recurrence time of $\sim$2.4 hrs. For our estimates of the Eddington ratio and black hole mass, the disk is expected be stable for $p_0 > 30$; assuming $\alpha \sim 0.1$, we find $p_0 \approx 1000 \gg 30$, meaning the disk is expected to be stable. Unstable solutions require unphysically high values of $\alpha \geq 100$, inconsistent with the lack of long-term evolution of the quiescent emission of the source \citep{Arcodia2024qpe2}.

Furthermore, \citet{Sniegowska23} find that magnetic instabilities require accretion disk sizes truncated to $\sim$10s of R$_g$ to reproduce the right timescales, which is inconsistent with our estimate of $R_{\rm out}$. 
Finally, our estimate of the outer radius of the disk is likely too small for limit-cycle oscillations due to ionization instabilities \citep{Janiuk11}.

\section{Summary}
\label{sec:summary}
We report on coordinated, time-resolved X-ray (XMM-Newton) and far UV (HST/STIS) observations of the X-ray QPE source eRO-QPE2. The \hst observations constitute the most sensitive UV observations of a QPE source to date, and cover both the eruption and quiescent phases. We detect a FUV point source with a mean (extinction-corrected) luminosity of $L_{\rm FUV} = 5 \pm 3 \times 10^{41}$ erg s$^{-1}$. No statistically significant FUV variability is detected between the eruption and quiescent X-ray phases down to a level of 1.8$\times$10$^{-18}$ erg cm$^{-2}$ s$^{-1}$ \AA$^{-1}$\ (3 $\sigma$; this corresponds to an extinction-corrected luminosity of 1--5$\times$10$^{40}$ erg s$^{-1}$, depending on the extinction correction used). 

Such a luminous FUV source cannot be explained by a compact nuclear stellar cluster with a young stellar population unless its parameters are extreme compared to known systems. We employ an accretion disk model to describe the X-ray to UV SED and find that this can explain the data if the accretion disk has a compact outer radius (\Rratio $= 86^{+36}_{-25}$, or \Rout $= 343^{+202}_{-138}$ R$_{\rm g}$ for a black hole mass of \Mbh = 5.9$\pm$0.3 M$_{\odot}$, as constrained from the model self-consistently). Such a compact accretion disk, much smaller than observed in AGNs, is a natural expectation following the tidal disruption of a star or unstable Roche-lobe overflow from a stellar companion.  
By accounting for the quiescent disk emission, we find that the ratio of X-ray to FUV luminosity of the eruption-only emission for any model is constrained to $L_{\rm peak, X}$ / $L_{\rm max, FUV} > 16-85$ (depending on the exact E(B--V) in the system).
With the results of this modeling in hand, we explore the implications for the various classes of theoretical models that have been proposed to explain QPEs.

Converting the QPE recurrence time to an orbital radius in the assumption of a quasi-circular orbit, we find that the putative orbiter is located at a distance that inevitably intersects with the compact accretion disk, making this scenario consistent with the class of object-disk collision models to explain the eruptions. 
Our modeling also allows us to constrain the Eddington-normalized accretion rate of the system, which is a parameter of interest in the class of accretion instability models to explain QPEs. We find $\lambda_{\rm Edd} = 0.13^{+0.18}_{-0.07}$, which in combination with the black hole mass and compact outer radius is inconsistent with the disk instability models that we considered.
Finally, we can also rule out QPE models that have either a classic, large AGN accretion disk, as well as those models where no accretion disk is present, to explain the QPE phenomenon in eRO-QPE2. \\

Future high spatial resolution observations at NUV and optical wavelengths can be used to more accurately constrain the extent of the accretion disk, because of the effect of disk truncation on the observed break in the UV/optical regime. 
A sample study using SED modeling including FUV and NUV wavelengths could be used to determine whether every known QPE source remains compatible with disk-object collision models while simultaneously constraining the black hole mass and Eddington ratio parameter space to test accretion disk instability models.

\begin{acknowledgments}
TW and MG acknowledge helpful discussions with T. B{\"o}ker, and I. Linial for providing an analytical expression to describe the multi-wavelength SED of the star-disk interaction model. 
We thank the referee for thoughtful comments that improved the manuscript.
Support for this work was provided by NASA through grant GO-17447 (P.I. Wevers) from the Space Telescope Science Institute, which is operated by AURA, Inc., under NASA contract NAS5-26555.  MG is supported in part by NASA XMM-Newton grant 80NSSC24K1885. This work was supported by a Leverhulme Trust International Professorship grant [number LIP-202-014]. 
The HST data presented in this article were obtained from the Mikulski Archive for Space Telescopes (MAST) at the Space Telescope Science Institute. The specific observations analyzed can be accessed via \dataset[doi: 10.17909/qn96-rf30]{https://doi.org/10.17909/qn96-rf30}.
This research was supported in part by grant NSF PHY-2309135 to the Kavli Institute for Theoretical Physics (KITP). We thank the organizers of the KITP program: Towards a Physical Understanding of Tidal Disruption Events, where part of this work was performed. 
\end{acknowledgments}
\vspace{5mm}
\facilities{ESO/VLT, HST, XMM-Newton}

\software{astropy \citep{2013A&A...558A..33A,2018AJ....156..123A}, HEASoft \citep{Arnaud1996}, BXA \citep{bxa}, }

\bibliography{bib}
\bibliographystyle{aasjournal}

\appendix
\renewcommand{\thefigure}{A\arabic{figure}}

\setcounter{figure}{0}
\setcounter{table}{0}

\section{Hubble Space Telescope data reduction}
\subsection{Detector dark background correction}
\label{sec:glow}
The STIS FUV-MAMA dark background consists of a low background level and a glow region that varies with temperature and thermal history of the detector (see Sec. 7.5.2 of \citealt{2023stii.book...23M} for more details\footnote{https://hst-docs.stsci.edu/stisihb/chapter-7-feasibility-and-detector-performance/7-5-mama-operation-and-feasibility-considerations$\#$id-7.5MAMAOperationandFeasibilityConsiderations-FUV-MAMADarkCurrent}). We fit both effects simultaneously using dark structure present in our science observations.

As input to our dark model, we identified 777 post-SM4 STIS/FUV-MAMA dark observations with exposure times $>$600 s from \hst cycles 17-31 in programs 11390, 11857, 12415, 12776, 13146, 13549, 13995, 14430, 14834, 14973, 15562, 15751, 16353, 16560, 16961, and 17390 (PIs: Proffitt, Zheng, Cox, Lockwood).  These range from 2009-06-09 to 2024-07-27 and measure 49M dark counts over 282 hours of total exposure time and over a range of detector temperature conditions.  These observations were scaled to count rates and binned to 2x2 low-resolution pixels.

Following the procedures of \citet{Lockwood2020}, we trained a scikit-learn \citep{scikit-learn} pipeline with these data, consisting of (1) a {\tt RobustScaler} step with centering and scaling using the 25-75th interquartile range, and (2) a Principal Component Analysis (PCA) step with 6 retained components.  These steps were selected for their applicability in fitting the detector dark behavior over short and long timescales, as well as fidelity when undergoing an inverse transformation.

Science observations from program 17447 were then fit with this pipeline by binning and applying the previously determined centering and scaling, and then calculating the dot product with the PCA eigenvectors.  A model super-dark was then generated using the pipeline's inverse transform.  Since the science observations include foreground counts, the fitted super-dark was subtracted and positive portions of the residual were identified as source regions in the input data, which were patched with data from the fitted super-dark.  This process was iterated until most foreground signal was identified and a super-dark was fit that does not over-subtract.

The resulting super-dark was linearly upsampled (https://scipy-cookbook.readthedocs.io/items/Rebinning.html$\#$Example-3) to the STIS high-res format (2048x2048) expected by {\tt CALSTIS} and saved to a FITS file.  We modified the DARKFILE keyword in the RAW science files to use these super-darks and processed them through {\tt stistools.calstis.calstis()}.

\subsection{Further processing}
Following our custom detector glow correction, we further pre-process the data to mitigate the effects of hot pixels and cosmic rays. We stack the observations for each \hst orbit, and use an iterative sigma-clipping scheme to clean detector artefacts such as hot pixels and charge traps. Note that for visit 1, the galaxy nucleus is located behind the repeller wire and we do not mask pixels flagged by the {\tt calstis} pipeline; for visit 2 this is not the case and we implement an additional masking scheme based on the DQ flags provided by the pipeline. Our results do not change significantly when masking data-quality flagged pixels for visit 1.

\section{XMM-Newton data reduction}
X-ray observations were taken as part of the joint \hst and \xmm program, with OBS-IDs 0932590101 and 0932590201  using the European Photon Imaging Camera \citep[EPIC;][]{Struder2001} in full frame mode with the thin filter. The observation data files (ODFs) are reduced using the \xmm Standard Analysis Software \citep[SAS;][]{Gabriel_04}.
The raw data files are processed using the \texttt{epproc} task. 
Given the higher sensitivity of the pn instrument, we do not include the MOS1/2 data in our analysis. 
We follow the \xmm data analysis guide to check for background activity and generate ``good time intervals'' (GTIs), manually inspecting the background lightcurves in the 10--12 keV band. 
Using the \texttt{evselect} task, we only retain patterns that correspond to single and double events (\texttt{PATTERN<=4}). Source spectra are extracted using a region of $r_{\rm src} = 35^{\prime\prime}$ around the peak of the emission. 
Background spectra are extracted from a region of $r_{\rm bkg} =
108^{\prime\prime}$ located on the same detector. The ARFs and RMF files
are then generated using the \texttt{arfgen} and \texttt{rmfgen} tasks,
respectively. 

\section{Fitting methodology}
\label{sec:fit}
The SED modeling and analysis are performed using the Bayesian X-ray Analysis software (BXA) version 4.0.7 \citep{Buchner2014}, which integrates the nested sampling algorithm \texttt{UltraNest} \citep{Buchner2019} with the fitting environment \texttt{PyXspec}. In this Bayesian framework, a probability distribution function is obtained for each parameter. UV photometry is added to \texttt{PyXspec} (without extinction correction) using the ``ftflx2xsp" tool available in HEASoft v6.33.2 \citep{Heasarc2014}, which generates the response file for the fitting package.
While X-ray spectra can be fit using Poisson statistics (a.k.a Cash statistics in \texttt{XSPEC}) in their native instrumental binning, \texttt{XSPEC} does not support fitting UV/optical/IR data with Poisson statistics. Therefore, the X-ray spectra are binned using an `optimal binning' scheme \citep{Kaastra2016}, ensuring that each bin contains at least 10 counts, and the simultaneous X-ray + UV fit is performed using Gaussian statistics (a.k.a. $\chi^2$-statistics in \texttt{XSPEC}).

To model dust attenuation intrinsic to the host galaxy, we use the \texttt{reddenSF} \texttt{XSPEC} model \citep{Guolo2024disksed}, which employs the \cite{Calzetti2000} attenuation law from 2.20 $\mu m$ to 0.15 $\mu m$ and its extension down to 0.09 $\mu m$ as described in \citet{Reddy2016}. The free parameter of the \texttt{reddenSF} model is the color excess E(B--V).

In \texttt{XSPEC}, the model we employ is 
\texttt{phabs$\times$redden$\times$zashift(phabs$\times$reddenSF$\times$diskSED)}. The Galactic 
X-ray neutral gas absorption is fixed to the Galactic hydrogen equivalent column density, $N_{H, G} =1.6 \times 10^{20}$ cm$^{-2}$ \citep{HI4PI2016}, and the Galactic extinction is 
given by E(B--V)$_{G}$ = 0.015 \citep{Schlafly2011}. The three parameters 
of \texttt{diskSED} (\Rins, \Tp, and \Rratio) are free to vary. 
The intrinsic host galaxy column density ($N_H$) is a free parameter, but the intrinsic dust extinction E(B--V) can not be free as we only have one UV band. Therefore, we assume a Galactic-like gas-to-dust ratio of 100, leading to the conversion $N_H ({\rm cm}^{-2}) = 2.21 \times 10^{21} \times A_V ({\rm mag})$ \citep{Guver09}. 

The marginalized posterior distributions of the fitting are shown in Figure \ref{fig:qpe2_post}, and the best-fit values for the model parameters can be found in Table \ref{tab:fitresults}.

\begin{figure*}[h]
	\centering
	\includegraphics[width=0.5\columnwidth]{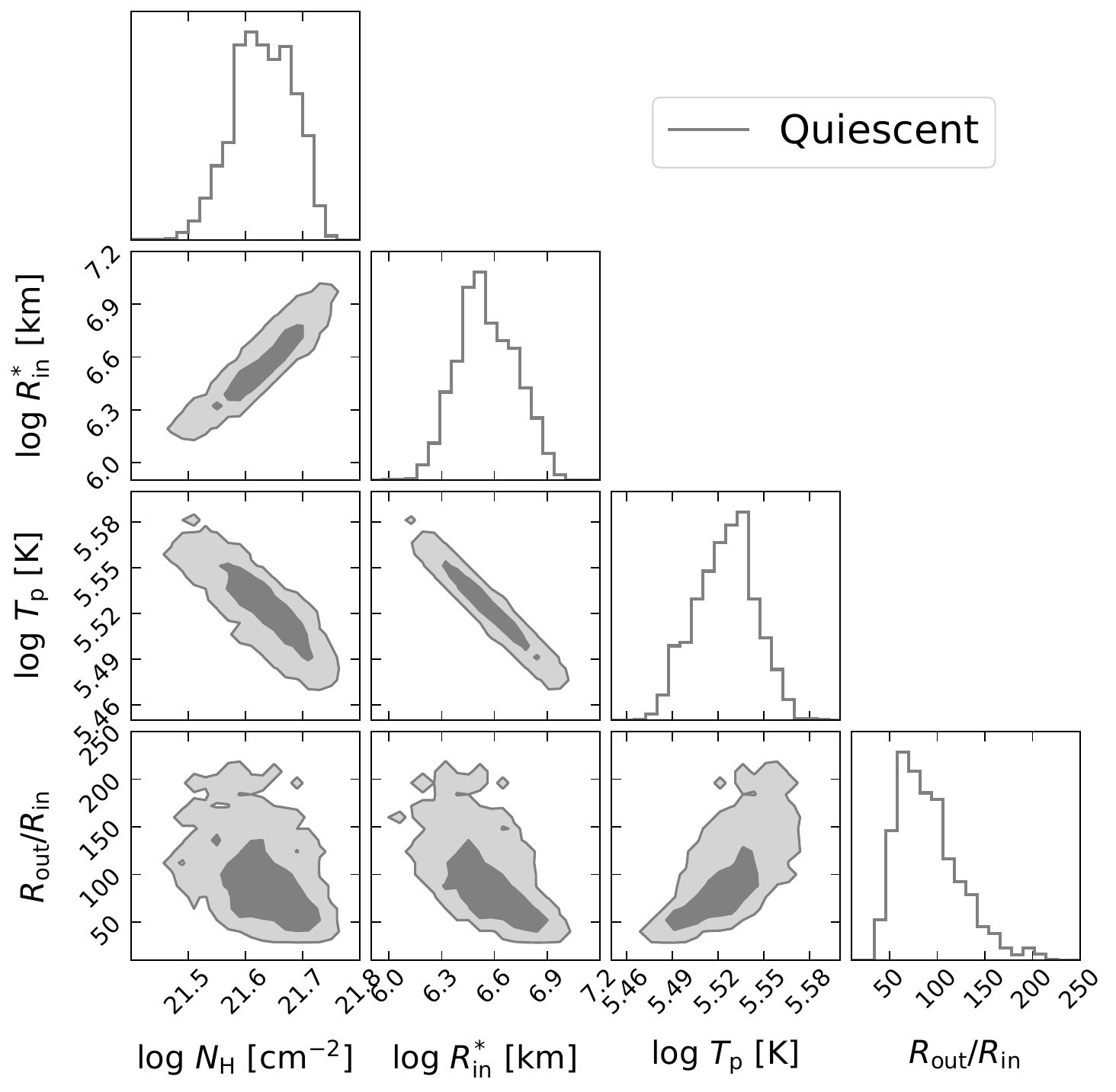}
    \includegraphics[width=0.4\columnwidth]{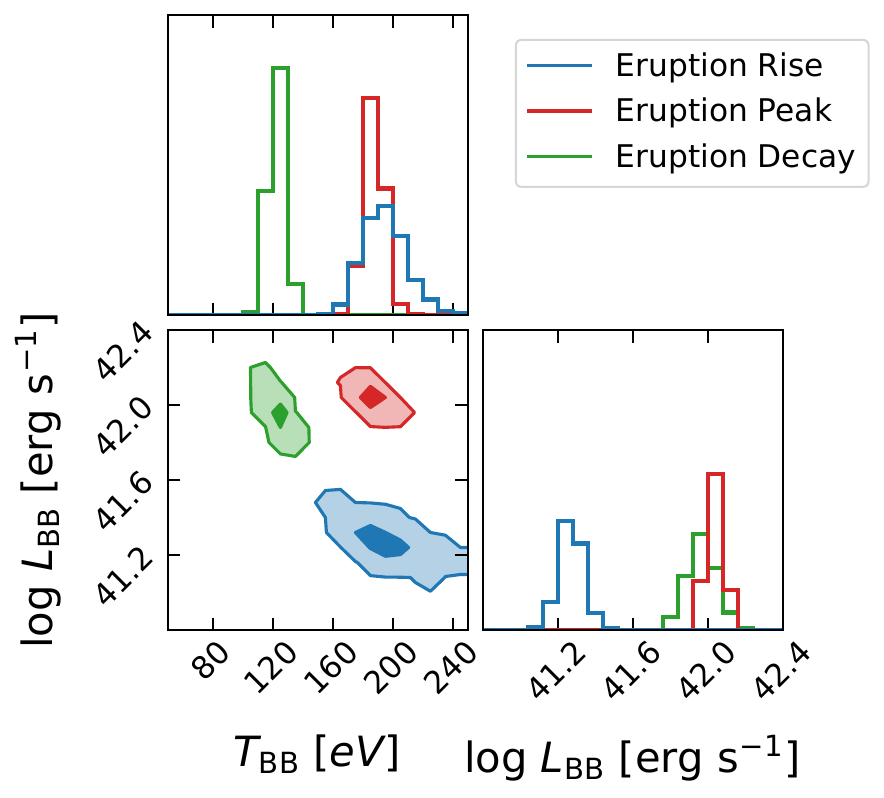}
	\caption{{\bf Full marginalized posteriors for the \texttt{diskSED} fits.} The left panel shows the quiescent component fit, while the right panel shows the eruption component phase-resolved fits. The color coding is the same as Figure \ref{fig:eruption_sed}). In the 2D histograms, the contours shows 68\% and 99\% of the probability distributions. }
    \label{fig:qpe2_post}
\end{figure*}

\begin{table}
    \centering
    \begin{tabular}{c|ccccccc}
        Phase  & Counts & log$_{10}$($N_H$) & log$_{10}$($T_p$) & log$_{10}$($R^{*}_{\rm in}$) & R$_{\rm out}$/R$_{\rm in}$ & $T_{BB}$ & log$_{10}$($L_{BB}$) \\
         & & (cm$^{-2}$) & (K) & (km) & & (eV) & (erg s$^{-1}$) \\\hline
        Quiescent & 782  & 21.63$_{-0.05}^{+0.06}$ & 5.52$^{+0.02}_{-0.02}$ & 6.54$_{-0.14}^{+0.19}$ & 86$^{+36}_{-25}$ & --- & --- \\
        QPE Rise &  223 & \vdots & \vdots & \vdots & \vdots & 193$\pm$ 15 & 41.3$\pm$0.1\\
        QPE Peak & 1151 & \vdots & \vdots & \vdots & \vdots & 187$\pm$ 6 & 42.1$\pm$0.1 \\
        QPE Decay & 522& \vdots & \vdots & \vdots & \vdots & 122$\pm$ 7 & 41.9$\pm$0.1 \\
    \end{tabular}
    \caption{Results of the joint fitting of the phase-resolved X-ray and UV SEDs. The eruption phase-resolved models differ only in the addition of a variable thermal component. Values and uncertainties denote the median and 68\% confidence interval.}
    \label{tab:fitresults}
\end{table}

\section{Black hole mass estimates}
\label{sec:mbh}
In addition to the black hole mass estimate based on the disk modeling (\Mbh = 5.9$\pm$0.3), two independent black hole mass estimates are available for eRO-QPE2. 
\citep{Wevers22} measured the central velocity dispersion in long-slit spectroscopy to be $\sigma_{\star}$ = 36$\pm$3 km s$^{-1}$, and more recently \citet{Wevers24b} reported a consistent measurement of $\sigma_{\star}$ = 38$\pm$6 km s$^{-1}$ based on spatially resolved MUSE data. 
Depending on the M--$\sigma$ relation that is used, this translates into M$_{\rm BH}$$\sim$ 10$^5$ M$_{\odot}$ with typical uncertainties of 0.4--0.5 dex. It is worth noting that none of the frequently used relations is anchored at such low velocity dispersions, and there may be significant systematic but unquantified uncertainties.

Alternatively, \citet{Mummery2024} reported a scaling relation between the late-time/plateau luminosity of a TDE and the mass of the disrupting black hole. In the assumption that the bright FUV point source is indeed a compact accretion disk produced by a TDE, we can use this scaling relation to obtain an independent estimate of M$_{\rm BH}$. We extrapolate our FUV measurement to an NUV luminosity by assuming these bands are located on the Rayleigh-Jeans tail of disk emission, that is, $\nu$$L_{\nu} \propto \nu{^3}$. Accounting for the difference in frequency (1.9$\times$10$^{15}$ Hz compared to the nominal NUV frequency of 1$\times$10$^{15}$ Hz), we infer $L_{\rm NUV}$ = 1.3$\times$ 10$^{41}$ erg s$^{-1}$, which translates into M$_{\rm BH}$ = 10$^{5.6\pm 0.5}$ M$_{\odot}$ (using Eq. 56 in \citealt{Mummery2024}).

These independent estimates consistently indicate the presence of a low mass black hole. For self-consistency, and given the potential of systematics in host galaxy correlations, we adopt a value of log$_{10}$(M$_{\rm BH}$) = 5.9$\pm$0.3 obtained from the accretion disk modeling. 
Finally,the inferred black hole mass when marginalizing over the full range of black hole spins a = [-0.99, 0.99] is log$_{10}$(M$_{\rm BH}$) = 5.6$^{+0.4}_{-0.3}$, remains fully consistent within the uncertainties.

\end{document}